# A new four states high deflection low actuation voltage electrostatic MEMS switch for RF applications.

Renaud Robin[1,2], Salim Touati[1], Karim Segueni[1], Olivier Millet[1], Lionel Buchaillot[2]

[1]DelfMEMS SAS, 5 rue Héloïse Haute Borne bâtiment CIEL, 59650 Villeneuve d'Ascq, FRANCE, (+33) 8 754 639 16

[2]IEMN, ISEN Dpt, Silicon Microsystems Group, Cité Scientifique - Avenue Poincaré, BP 60069, 59652 Villeneuve d'Ascq Cedex, FRANCE, (+33) 3 201 979 79

***Abstract* — This paper presents a new electrostatic MEMS (MicroElectroMechanical System) based on a single high reliability totally free flexible membrane. Using four electrodes, this structure enables four states which allowed large deflections (4µm) with low actuation voltage (7,5V). This design presents also a good contact force and improves the restoring force of the structure. As an example of application, a Single Pole Double Throw (SPDT) for 24GHz applications, based on this design, has been simulated.**

## I. PRINCIPLE

The structure consists in a membrane of gold simply supported over three pillars. To complete the system, four actuation electrodes are used: two centered and one at each extremity of the membrane. These electrodes enable four states. The first one called rest state is obtained when no electrodes actuation are used, there is no deflection in this case. Two other states are obtained when either odd or even electrodes are actuated. A large positive deflection is then noticed on one side (up state) whereas on the other side a smaller negative deflection of the membrane is obtained (down state). Finally, another state can be show using the two external electrodes. It results in a very large deflection of the entire membrane which allowed to restore the membrane in rest state, even if stiction occurs, due to a lever effect. (Fig1.)

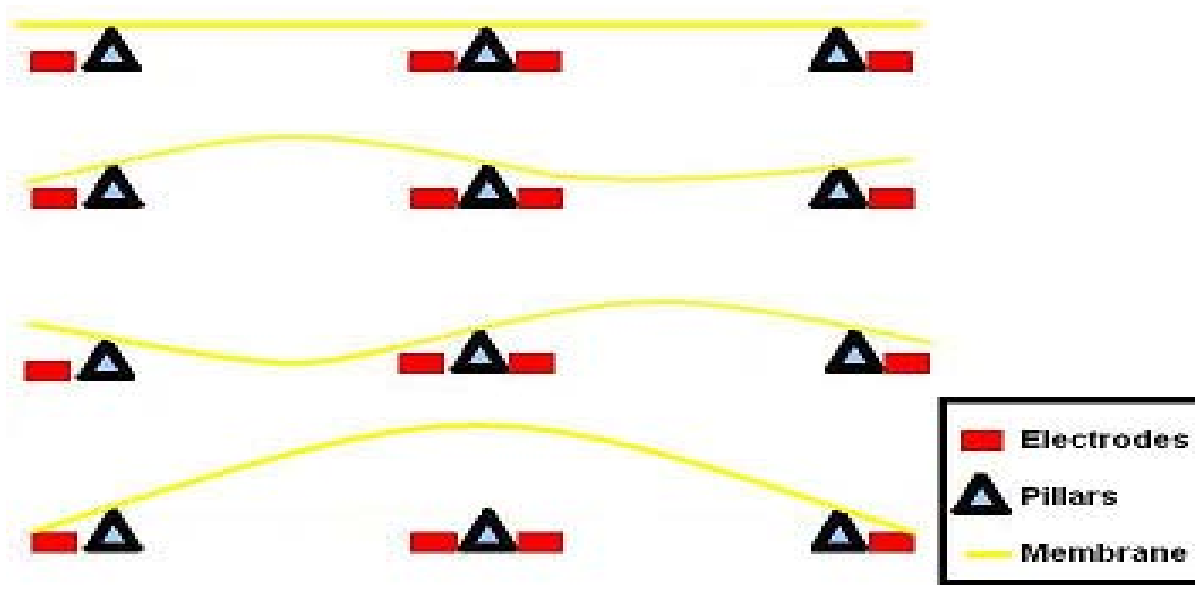

Fig1. Schematic view of the four actuation states.

The design of the structure has been performed to obtain such deflections at low actuation. That's why large area have been designed on both membrane and electrodes to maximize the actuation force. Wings of balancing are used to reduce stress concentration at the contacts structure-pillar and also to avoid zipping effect [1]. Small arms of correlation have been designed in order to transmit the movement of one side to another without limiting the deflections magnitude. These arms permit also to reduce the actuation voltage and to shift the area of maximal deflection. Finally, mechanical stops units have been used to maintain the structure without limiting the movements and the deflections when actuated. They are realized by auto-alignment to ensure the limited x-y-z translations. (Fig2.)

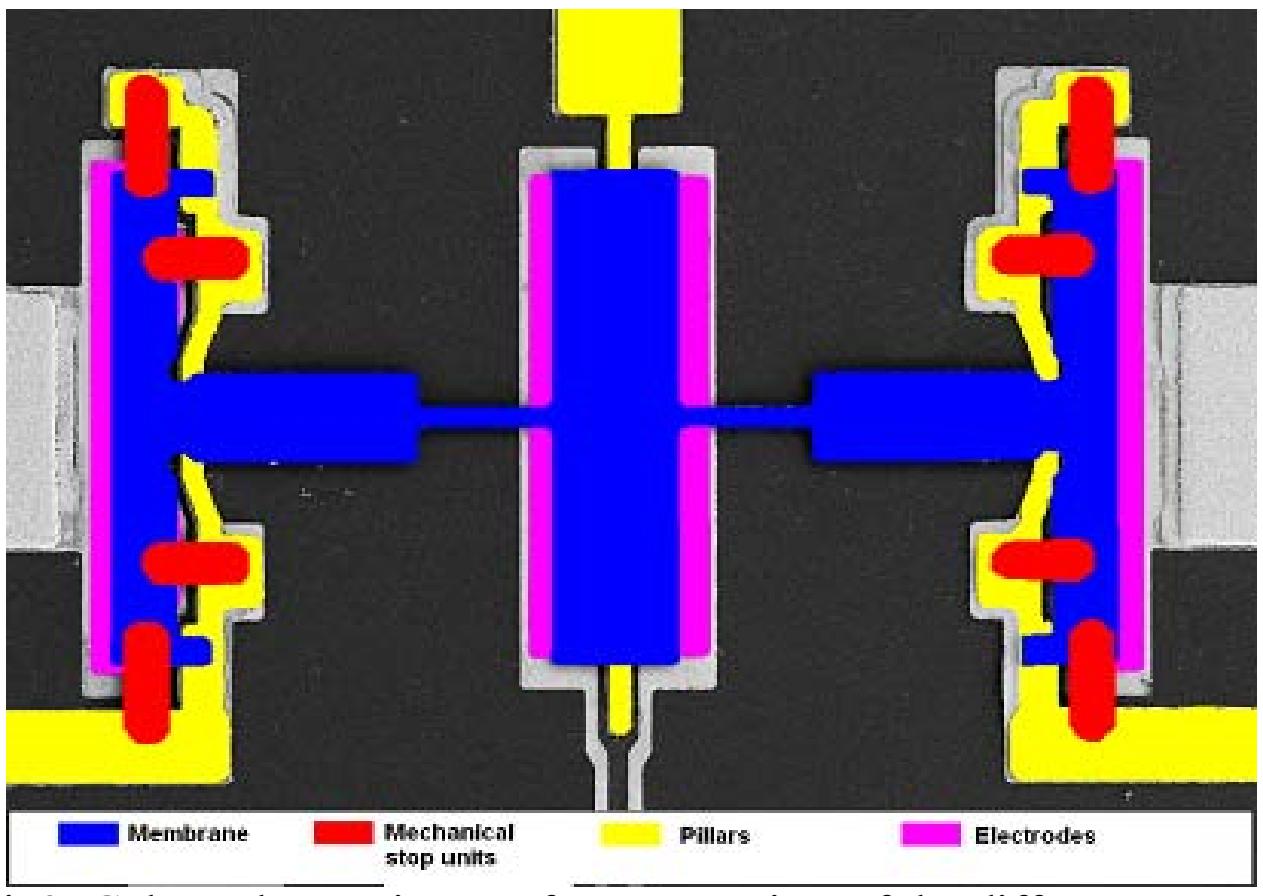

Fig2. Coloured sem picture of an upper view of the different parts of the structure.

## II. MECHANICAL COUPLED-FIELD ANALYSIS

The simulations have been done using sequentially coupled physics analysis. In order to have a good estimation of the voltage required, an air volume has been meshed rather than use Trans126 element to take into account the side effects. Contacts between membrane-electrodes or membrane-transmission line have been created using Targe170 and Conta173 elements. The membrane's material is a 2µm thick gold. The size of the structure is about 600µm long and 320µm large. The gap between the membrane and the electrodes is 3µm.

The maximal deflection of a simply supported structure is function of the ratio of the length between two pillars and the total length [2]. The use of small arms of correlation shows an increase of the deflection and a better transmission of the movement with induced stress divided by ten. The length of these arms affects where the deflection is maximal. With arms of 80µm long, deflections of 4µm have been demonstrated with an actuation voltage of only 7.5V. (Fig3.)

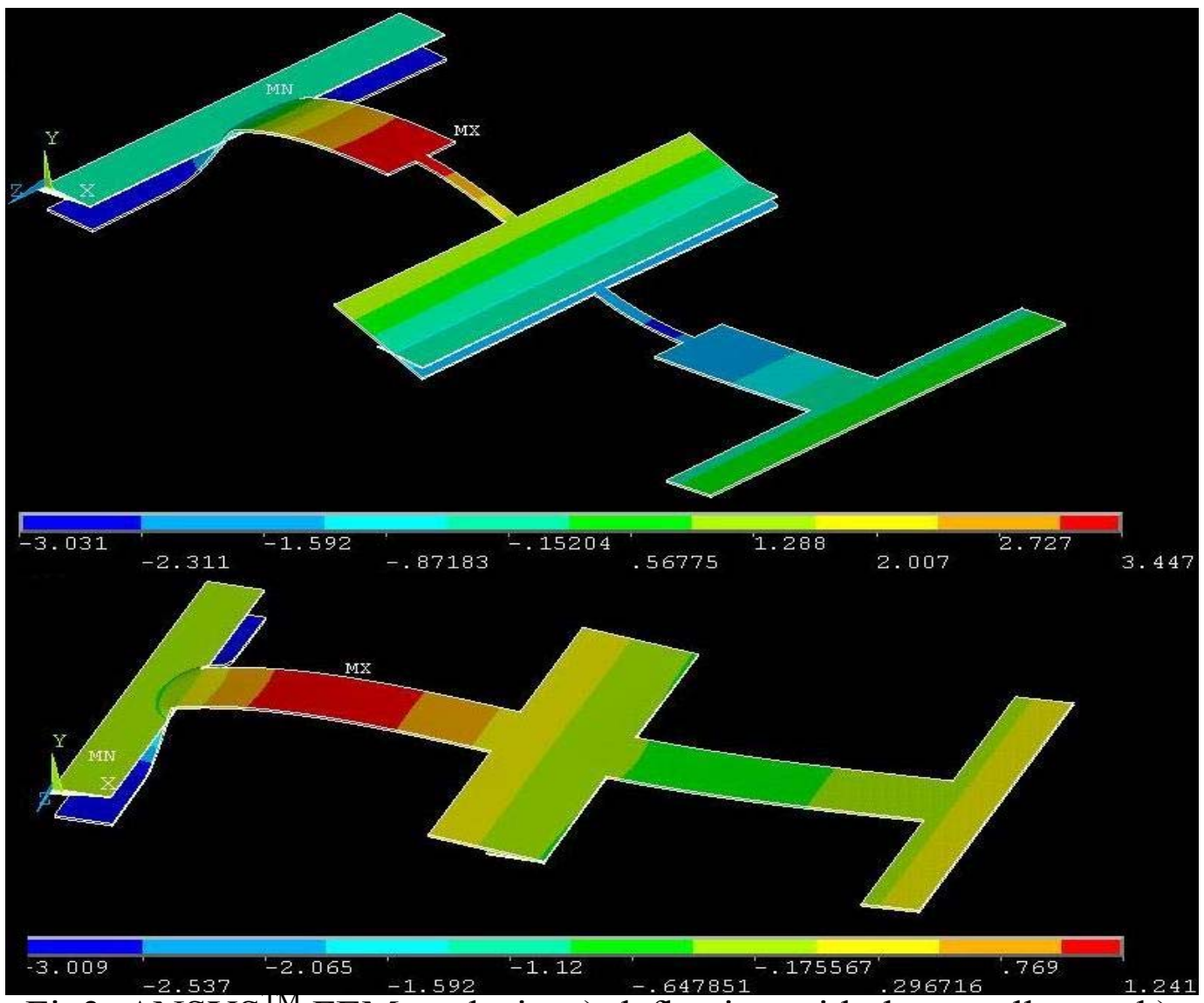

Fig3. ANSYS[TM] FEM analysis. a) deflection with the small arms b) deflection without the arms. Both are done with 7,5V actuation.

The fact both sides are correlated improve contact force between the transmission line and the membrane for the side in down state in case of a SPDT application [4]. The behavior of the structure is similar to a push pull design. The largest the deflection of one side is, the higher the contact forces on the other side are. Contact force of more than 10µN has been simulated. Moreover the switching time for such design is lower [3] because the stiffness is greater than for a single structure. The simulated resonant frequency of the structure is 23 kHz. The wings at each extremity of the membrane cause four major effects: the actuation voltage is a bit raised and the stress level is reduced; the zipping effect is limited; and finally the harmonic frequencies are higher than the fundamental which suppress spurious modes (Fig4.).

The increase of the actuation voltage is caused by the stiffness added by these wings. The reduced zipping effect is a consequence of a more homogeneous displacement of the membrane. The self actuation is also avoided with this design because the large deflections combined with the stiffness prevent risk to return in rest state when a signal is transmitted in the line. More over, because of the degrees of freedom used, there is no dependence in temperature with such kinds of structure. The gap between the membrane and the mechanical stop units is 800nm. It has been simulated that this gap allows an elevation of temperature of more than 100°C without created contacts between the structure and the stop units. The behavior of the structure is still the same when the temperature is raised which also improve the reliability of this design.

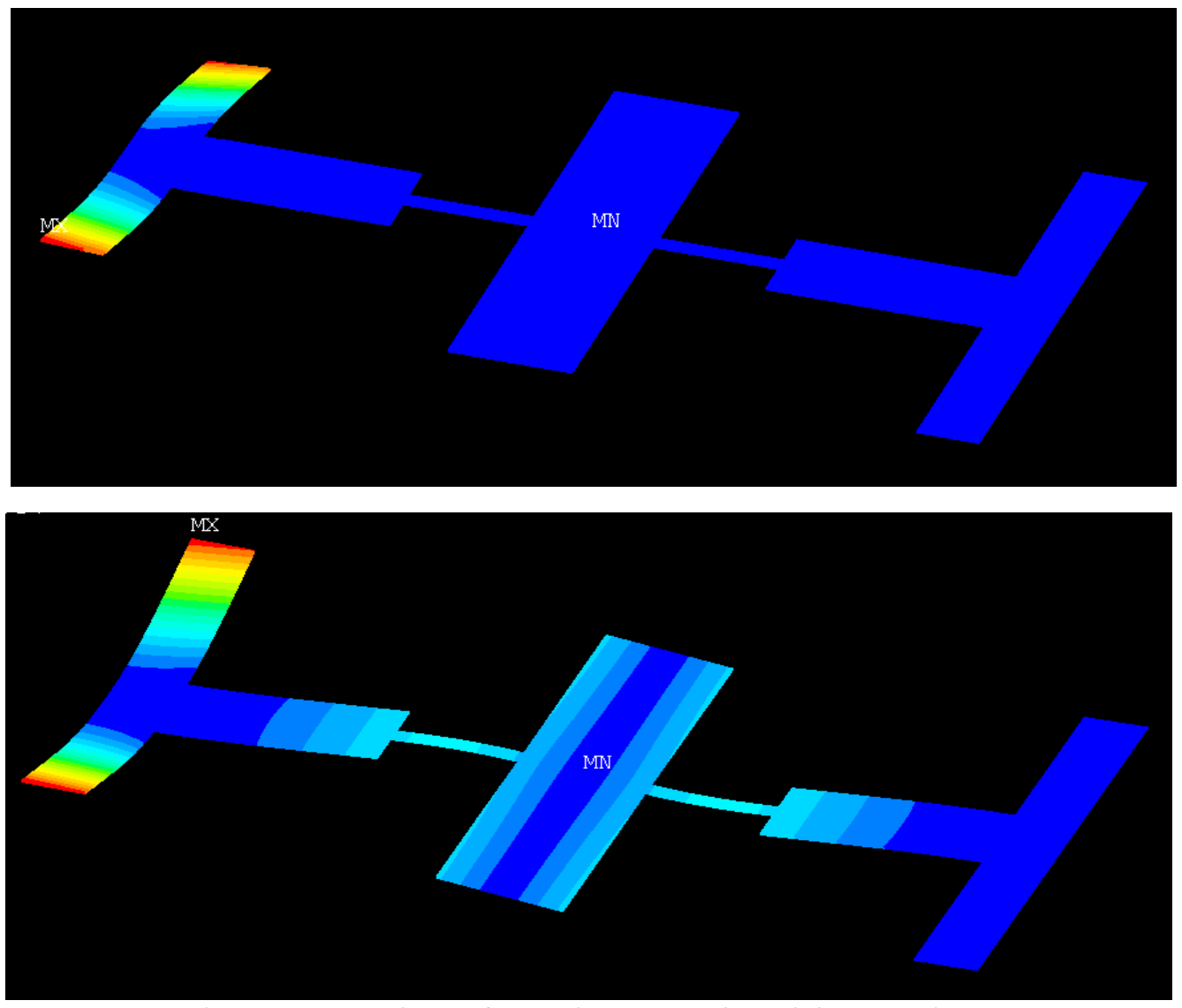

Fig4. Examples of spurious mode without wings.

III. PROCESS FLOW AND MECHANICAL CHARACTERIZATION

A low temperature process (<200°C) is used, based on surface micromachining, and combined with two sacrificial layers of chromium and silicon dioxide. Six masks are needed for this process. The first step consists in patterning 200 nm actuation gold electrodes and 100 nm PECVD (Plasma Enhanced Chemical Vapor Deposition) Si3N4 layer for the dielectric insulation. Then a 200 nm Chromium layer is sputtered in order to protect the Si3N4 during releasing step. This chromium layer is also used for electrical interconnection for the next gold electroplating. A 2.8µm SiO2 is then deposited patterned and etched by plasma etching; this layer is used as mold for electroplating of pillars. A 300 nm thick chromium layer is sputtered to avoid contact between the membrane and the pillars. The structural membrane is performed by a 2 µm thick electroplated gold. A third 500nm chromium layer is deposited and the mechanical stop units are patterned by a gold electroplating in a positive photoresist mold. Finally releasing consists in etching silicon dioxide in hydrofluoric acid (HF) and chromium in chromium etchant before drying the structures using super critical $CO_2$ dryer.

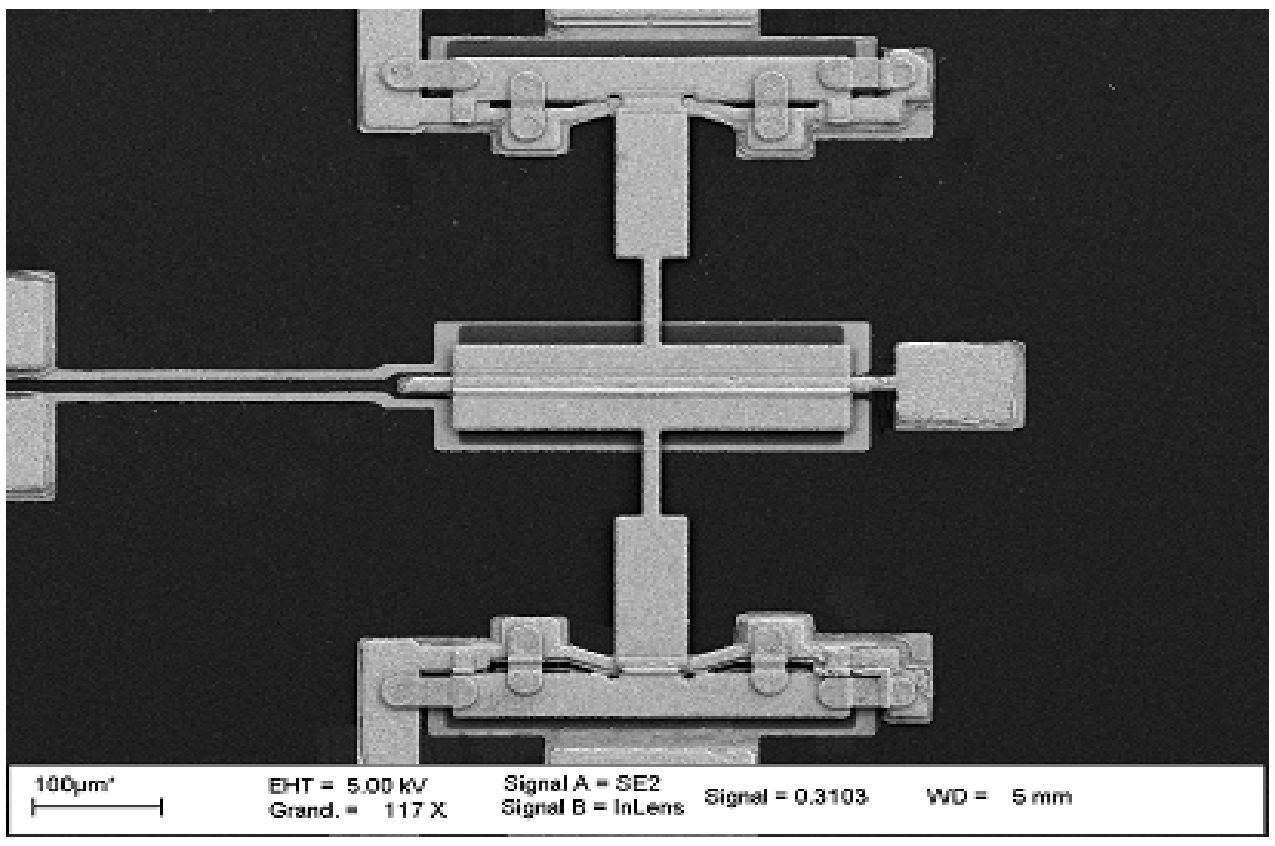

Fig5. Sem view of the structure after releasing and drying step.

Fabricated structures have been actued under probe station. A unipolar rectangular 2Hz electrical signal has been applied between the membrane and the electrodes (Fig6.). More than twenty structures have been actuated. Some were actuated in fourth state only and others in odd or even state, and finally a few were actuated in all states. It allows to compare the mechanical behaviour of structures actuated in only one state and others actuated in different states.

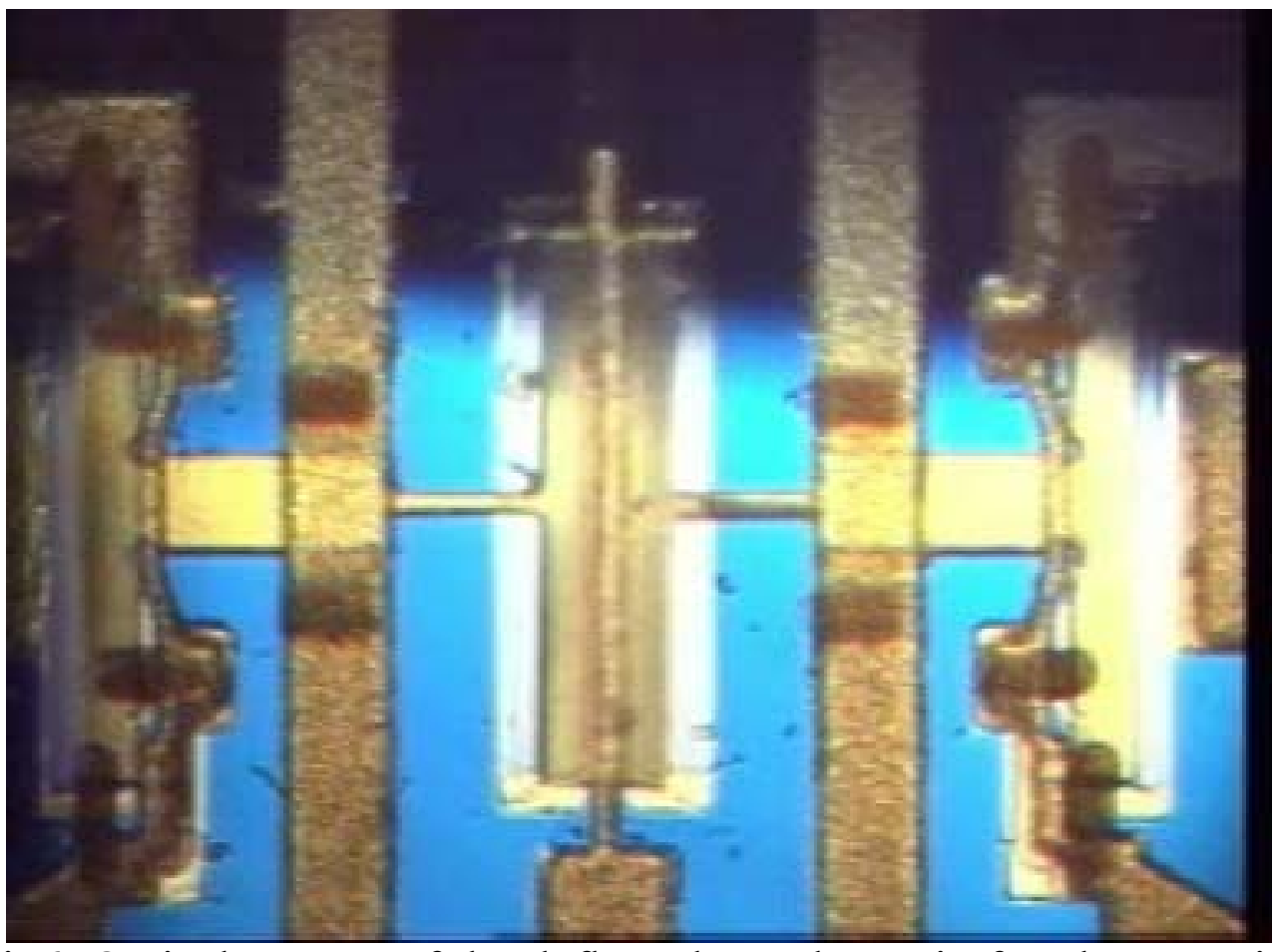

Fig6. Optical capture of the deflected membrane in fourth state with MS lines.

No difference were noticed and the mechanical behavior was assumed to be the same in each case thanks to the degree of freedom of the structure. The minimal pull-in voltage was estimated at 17V in case of external actuation (fourth state) and 19V for odd or even actuation. Fig7. shows the value of actuation voltage at pull-in in fourth and odd or even state. Blue squares stand for odd or even mode and red triangles for fourth state. As said before it is not necessary the same structures actuated in both mode. Without Radio-Frequency (RF) characterization it is not possible to determine precisely the pull-in voltage value. The value indicated here correspond to large excursion optically noticed thus the real pull-in value may be inferior. The higher value of pull-in in odd or even mode can be explain by this. The excursion in fourth mode is greater thatswhy it can be noticed at lower voltage actuation.

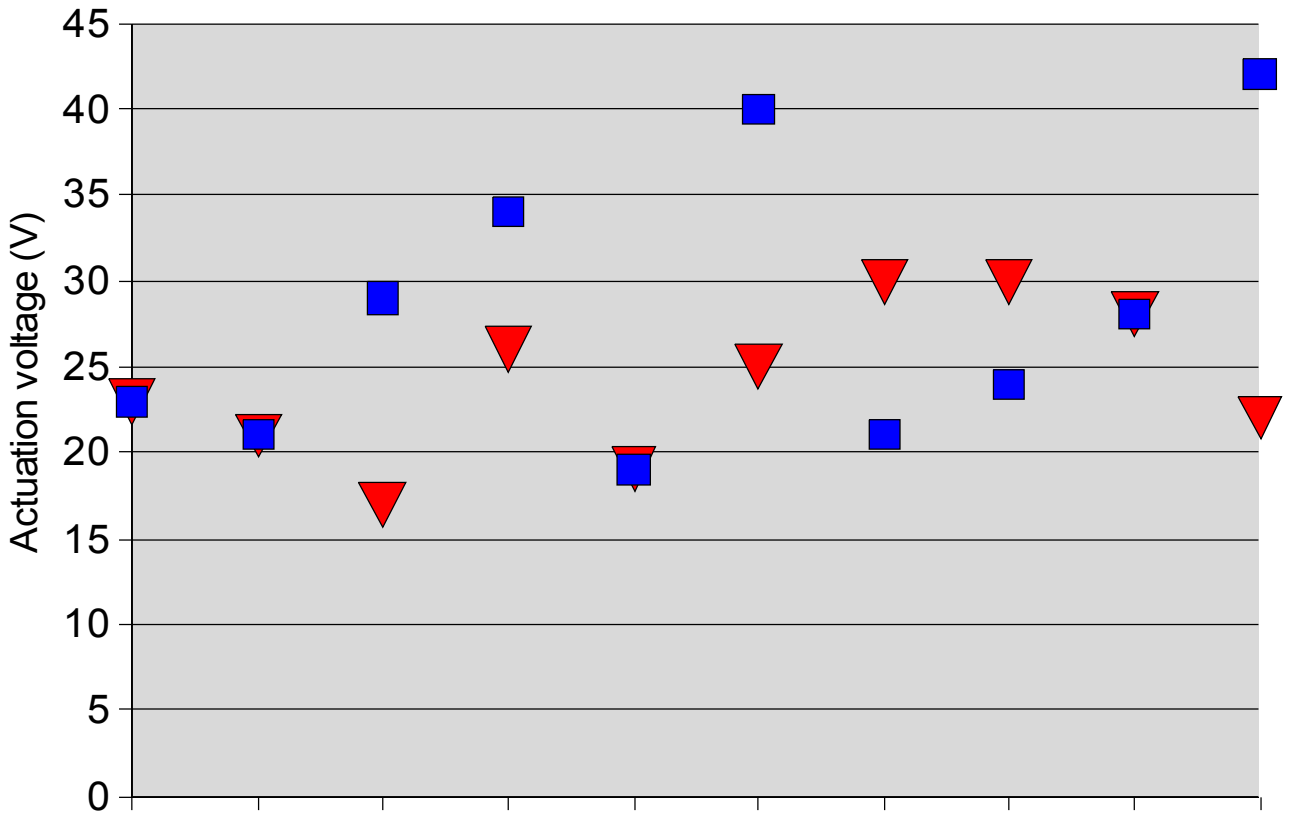

Fig7. Dispersion of the pull-in voltage value in different mode.

The difference between simulations and measurements is due to the fact that the gap was larger than the simulated gap on one hand and that the membrane was deformed on other hand. For the first point, the gap has been measured by interferrometrical microscope and it has been shown that the real gap was about 3,4µm. (Fig8.)

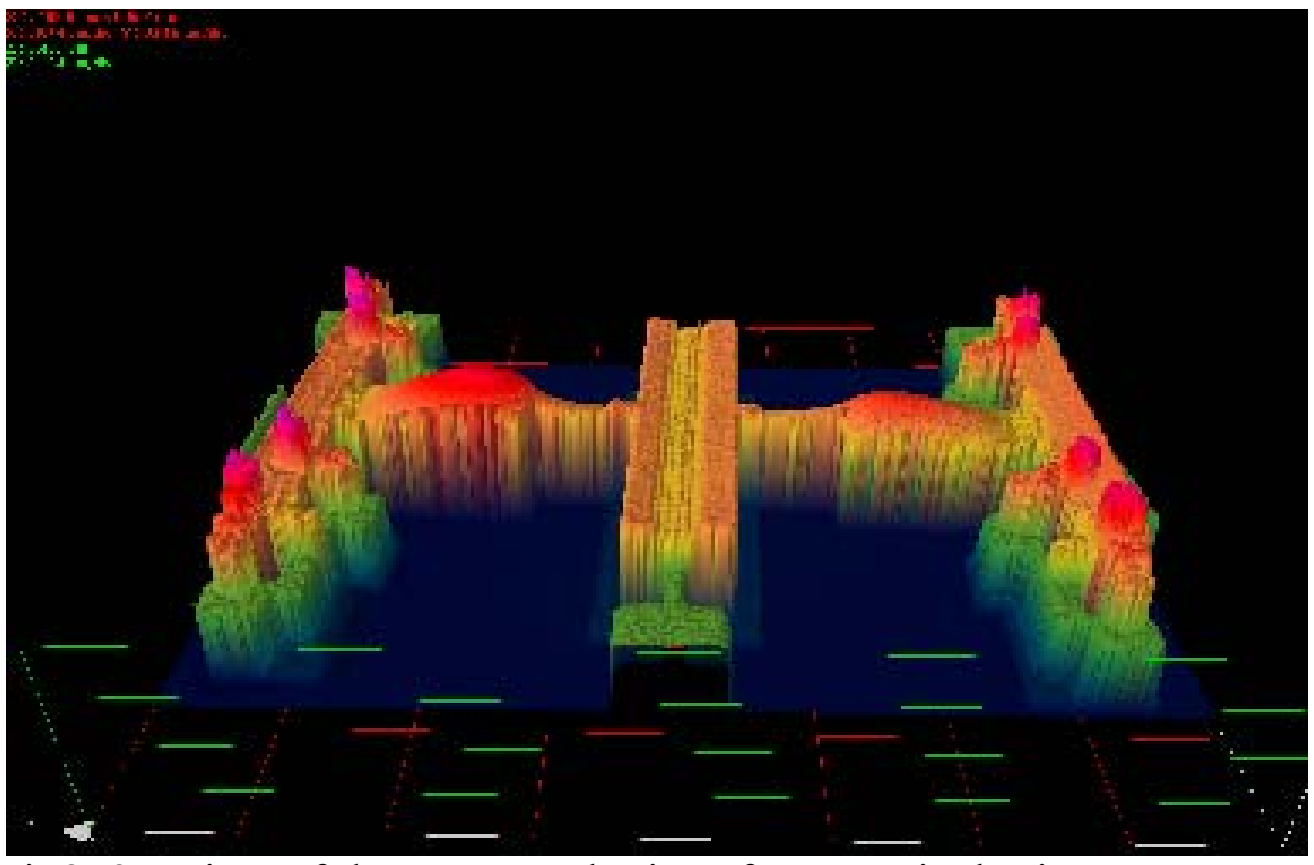

Fig8. 3D view of the structure by interferrometrical microscope.

It also demonstrates that the membrane was deformed by stress caused by unwanted electroplated gold which rigidified locally the structure during the creation of the mechanical stop units. The shape of the structure is not plane, it looks like a buckling mode. In this spurious mode the gap is bigger where the electrodes are located which means higher actuation voltage value. Retrosimulations with measured gap value and a 50MPa compressive stress on the membrane shows the same order of actuation voltage. To reduce the pull-in voltage and fit better to the original simulations, the electrolysis step had to be better controlled.

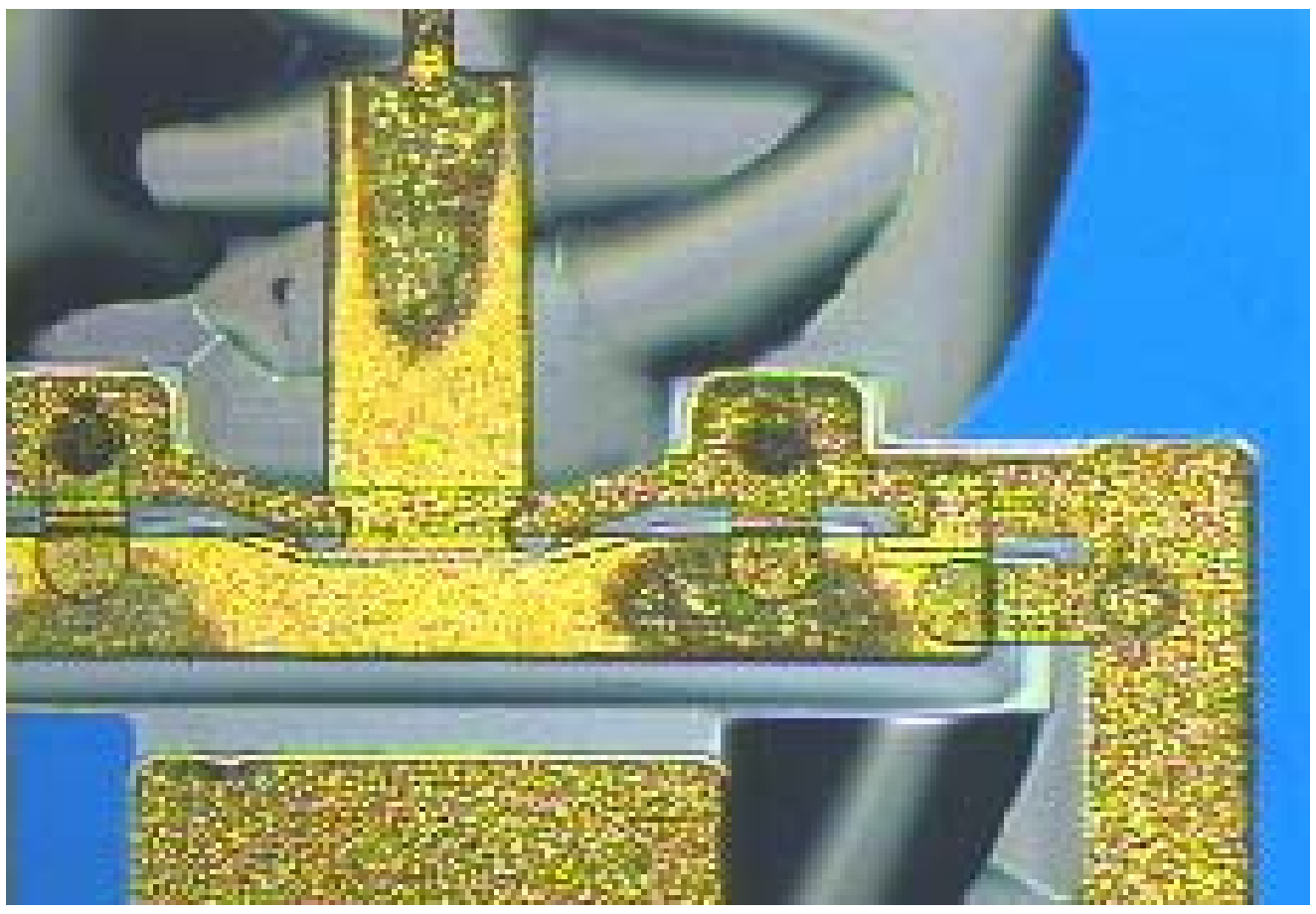

Fig9. Optical view of the unwanted electroplated areas during releasing step. (chromium etching)

Mechanical characterization also demonstrated that when the membrane is sticking in odd or even state by dielectric charging, the external electrodes actuation allow to unglue the membrane and restore the membrane in rest state[5]. This is possible because the restoring force is combined with the actuation force of the fourth state which are both opposed to the sticking force. With such design, attractive forces caused by stiction or high power under the membrane can be ignored because the restoration state can be obtained with the fourth mode actuation.

IV. APPLICATION TO A 24GHZ SPDT

With the odd and even mode, a SPDT application can be achieved with such design using one structure and two transmission lines. This RF design uses microstrip lines to avoid coupling effects which can occurs between the two outputs with coplanar waveguide. The width of the lines is 60µm and the substrate is high resistivity silicon with a permittivity of 11.9. The simulations have been done using HFSS™ 10.1. The structure consists of a double capacitive shunt switch with three microstrip lines, one for input and two others for the outputs designed for 24GHz applications. (Fig10.) The switch is the same structure as described before with a PECVD silicon nitride layer for the capacitive contact [4].

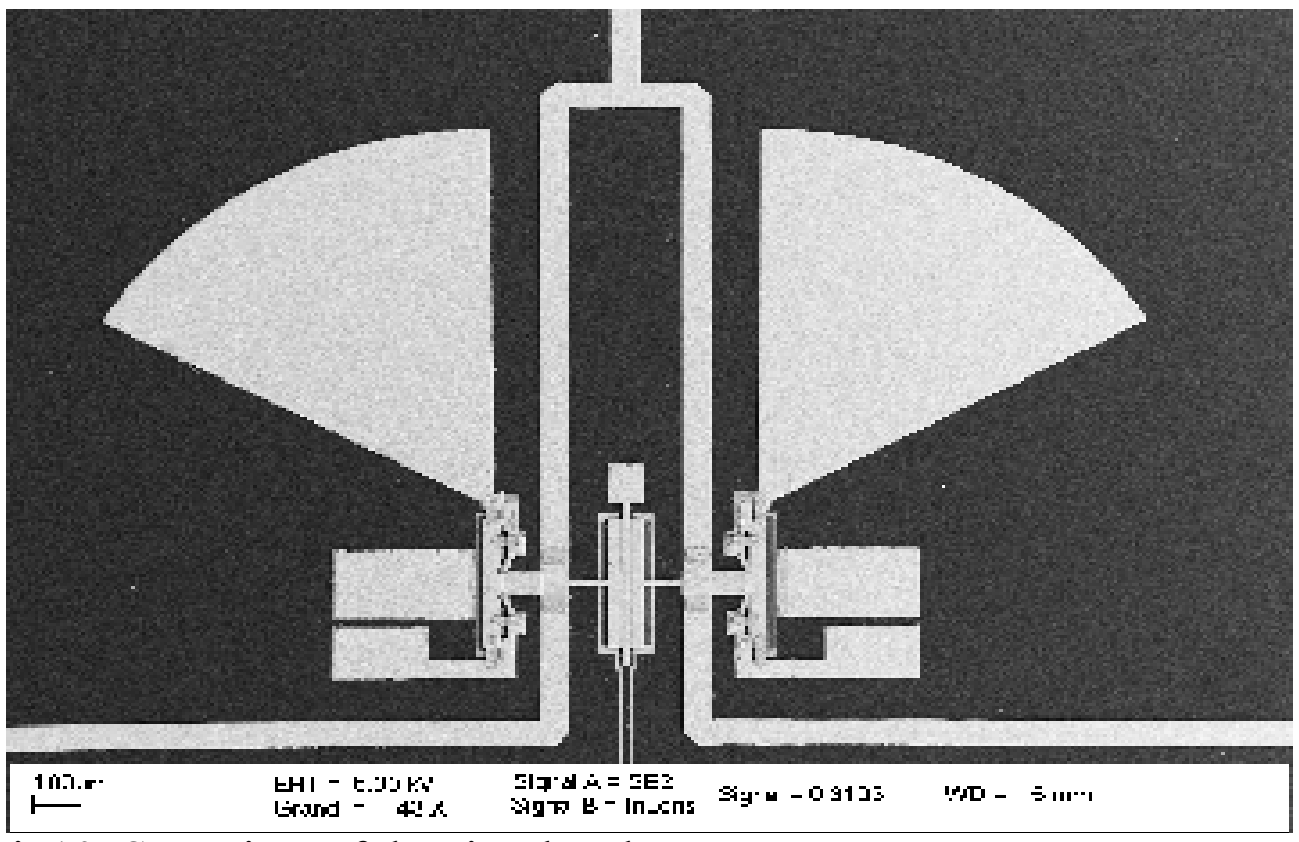

Fig10. Sem view of the simulated structure.

When in up state, the large deflections permit low return losses of 28dB at 24GHz. (Fig11.) In down state, the push pull generates large contact forces which create a bigger capacity. The switch exhibits isolation between the input and output higher than 30dB at 24GHz because the membrane is slapped on the microstrip line due to the lever effect. The isolation between the two outputs is a bit lower: 25dB at 24GHz. The insertion losses at such frequency are 0.65dB for 2mm line's length.
Radial stubs have been designed to create a virtual ground to ensure the shunt functionality. The choice of radial stubs rather than vias has been done to have an easier fabrication process even if radial stubs works for a smaller bandwidth. Coupling between stubs and transmission line is used to improve the adaptation level. A quarter wave line is placed between the capacitive contact and the microstrip separation to transform the short circuit in open circuit and thereby presents an infinite impedance for the blocked line.

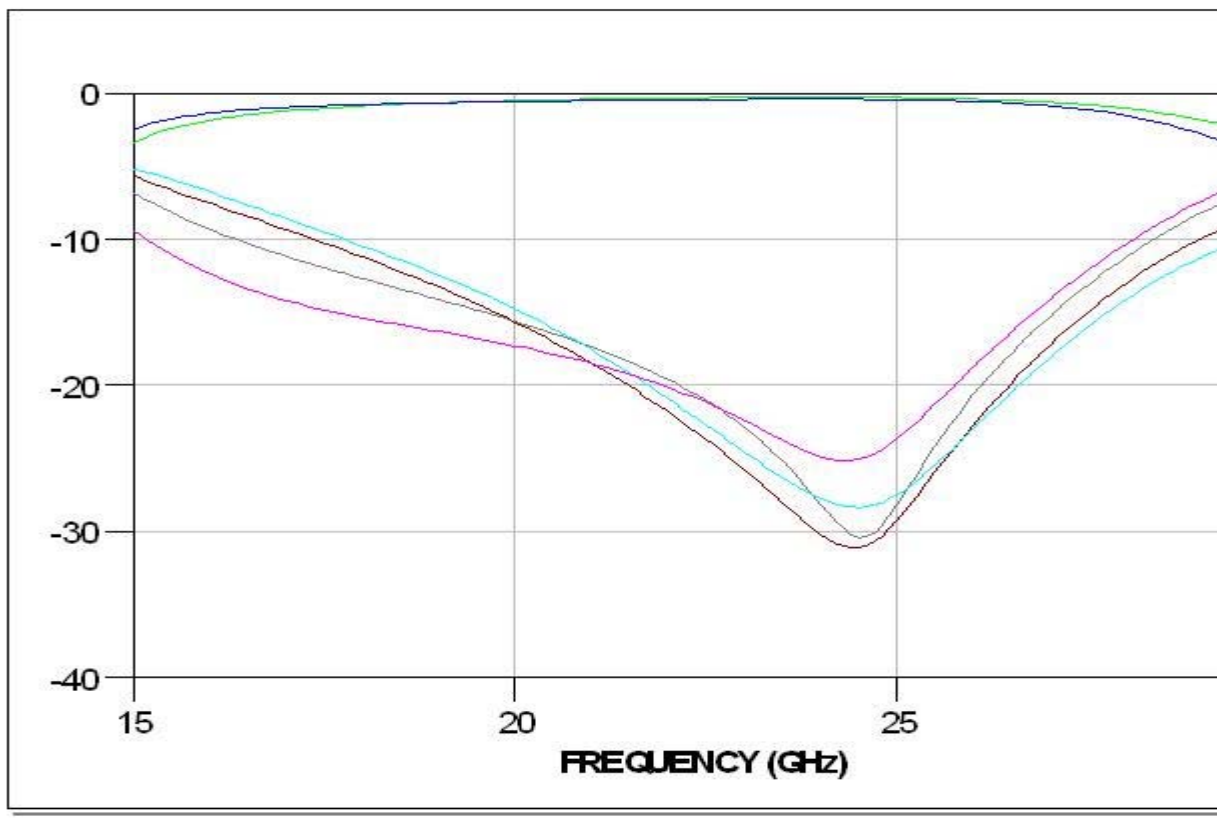

Fig11. Simulated S parameters of the SP2T.

V. CONCLUSION

An innovative MEMS switch showing high reliability mechanical design with low actuation voltage has been demonstrated in this paper. The push pull principle combined with free flexible membrane increase mechanical performances and prevents some failure problems of others RF MEMS switches like self actuation or low contact forces. Fabricated structures have been mechanically characterized and low actuation pull-in voltage have been noticed (around 20V), showing good agreements with the simulations. The actuation voltage can be reduced if the stress generated by the different electrolysis steps is better controlled. A SPDT application has already been simulated with this principle. It exhibits good RF performances at the focused frequency (24GHz). RF characterizations of micromachined SPDT are expected for the next year with lower pull-in voltage.